\documentclass{aa}
\usepackage{graphicx}
\usepackage{longtable}
\usepackage{latexsym}
\usepackage{amssymb}
\usepackage{lscape}
\input{psfig.txt}
\begin{document}
\title{H$\alpha$ surface photometry of galaxies in the Virgo cluster. IV: the current star
formation in nearby clusters of galaxies
\thanks{based on observations taken at the Observatorio Astron\'omico Nacional (Mexico),
the OHP (France), Calar Alto and NOT (Spain) observatories.}
}

\author{G. Gavazzi \inst{1}
\and A. Boselli \inst{2}
\and P. Pedotti \inst{1}
\and A. Gallazzi \inst{1}
\and L. Carrasco \inst{3,4}
}

\authorrunning{G. Gavazzi}
\titlerunning{H$\alpha$ surface photometry of galaxies in the Virgo cluster}

\offprints{G. Gavazzi}
\institute{
Universit\`a degli Studi di Milano-Bicocca, Piazza delle scienze 3, 20126 Milano, Italy\\
\email {giuseppe.gavazzi@mib.infn.it}
\and
Laboratoire d'Astrophysique de Marseille, Traverse du Siphon, F-13376 Marseille
Cedex 12, France\\
\email {Alessandro.Boselli@astrsp-mrs.fr}
\and
Instituto Nacional de Astrof\' \i sica, Optica y Electr\'onica,
Apartado Postal 51. C.P. 72000 Puebla, Pue., M\'exico\\
\email {carrasco@transun.inaoep.mx}
\and
Observatorio Astron\'omico Nacional, UNAM, Apartado Postal 877, C.P. 22860, Ensenada B.C., M\'exico
}
\date{}

\abstract{
H$\alpha$+[NII] imaging observations of 369 late-type (spiral) galaxies in the Virgo cluster and in 
the Coma/A1367 supercluster are analyzed, covering 3 rich nearby clusters (A1367, Coma and Virgo)
and nearly isolated galaxies in the Great-Wall. 
They constitute an optically selected sample ($m_p<16.0$) observed with $\sim 60 \%$ completeness.
These observations provide us with the current ($T<10^7$ yrs) star formation
properties of galaxies that we study as a function of 
the clustercentric projected distances ($\Theta$). The expected decrease of the star formation rate (SFR),
as traced by the H$\alpha$ E.W.,  
with decreasing $\Theta$ is found only when galaxies brighter than $M_p \sim -19.5$ are considered. Fainter objects 
show no or reverse trends. 
We also include in our analysis Near Infrared data, providing us with 
informations on the old ($T>10^9$ yrs)
stars. Put together, the young and the old stellar indicators give the ratio of 
currently formed stars over the stars formed in the past, or "birthrate" parameter $b$. For the considered galaxies
we also determine the "global gas content" combining HI with CO observations. We define the
"gas deficiency" parameter as the logarithmic difference between the gas content of isolated galaxies
of a given Hubble type and the measured gas content.
For the isolated objects we find that $b$ decreases with increasing NIR luminosity. 
In other words less massive galaxies are currently forming stars at higher rate than 
their giant counterparts which experienced most of their star formation activity at earlier
cosmological epochs.
The gas-deficient objects, primarily members to the Virgo cluster, have their birthrate significantly lower than
the isolated objects with normal gas content and of similar NIR luminosity. 
This indicates that the current star formation is regulated by the gaseous content of spirals.
Whatever mechanism (most plausibly ram-pressure stripping) is responsible for the pattern of gas 
deficiency observed in spiral galaxies members to rich clusters, it also produces the observed quenching of the
current star formation.
A significant fraction of gas "healthy" (i.e. with a gas deficiency parameter less than 0.4)
and currently star forming galaxies
is unexpectedly found projected near the center of the Virgo cluster. Their average Tully-Fisher distance
is found approximately one magnitude further away ($\mu_o=$31.77) than the distance of their gas-deficient counterparts
($\mu_o=$30.85), suggesting that the gas healthy objects belong to a cloud projected onto the cluster center, 
but in fact lying few Mpc behind Virgo, thus unaffected by the dense IGM of the cluster.
\keywords{Galaxies: Galaxies: photometry; Galaxies: clusters: individual: Virgo}
}

\maketitle

%

\section{Introduction}
A significant trend of the global star formation rate (SFR) of galaxies with the projected 
clustercentric distance from rich clusters of galaxies is well documented in the local universe 
($0.05<z<0.1$). The mean SFR, as traced by the equivalent width of the H$\alpha$ line (Kennicutt 1989), 
is found to decrease with decreasing distance from rich clusters (Lewis et al. 2002). 
This pattern is dominated by the "morphology segregation" effect (Dressler 1980), i.e. 
elliptical and spheroidal galaxies with little or no current star formation overcome in 
number the star forming galaxies in the center of rich clusters. What physical mechanism 
(nature vs. nurture) is responsible for the morphological transformation taking place in 
the densest environments is however not yet fully understood. To shed light on the various 
possibilities, i.e. galaxy harassment (Moore et al., 1996, 1998), tidal stirring (Mayer et al. 2001) 
or ram pressure stripping (Gunn \& Gott 1972), it is crucial to establish observationally if,
beside the morphology segregation, 
galaxies of a given morphological type, namely the spirals, are affected by a systematic SFR decrease 
toward the center of nearby clusters. \\
If on one hand Kennicutt (1983) found that spirals in the Virgo cluster
have their mean SFR as much as a factor of two lower than isolated galaxies,
Gavazzi et al. (1998) did not confirm this evidence in the Coma and A1367 clusters.
Moreover Iglesias-Paramo et al. (2002) found that the shape of the H$\alpha$
luminosity function of these two clusters does not differ significantly from the one
of isolated galaxies.
The result of Kennicutt (1983) was based on only a dozen giant galaxies with
H$\alpha$ measurements from aperture photometry, 
thus requiring a confirmation on a larger sample with modern imaging data. \\
With the aim of solving this riddle we undertook an H$\alpha$ imaging survey
of two optically complete samples of galaxies. 
The first is composed of nearly isolated objects selected from the CGCG 
(Zwicky et al. 1961-68) in the bridge between Coma and A1367,
which we observed down to the limit of 15.7 mag. This constitutes our reference sample of non-cluster objects. 
The cluster sample is focused on A1367, the Coma and the Virgo clusters. 
We took H$\alpha$ imaging observations of these regions (Gavazzi et al. 1998, Gavazzi et al.
2002a, Paper I of this series; Boselli \& Gavazzi 2002; paper II and  
Boselli et al. 2002b; paper III). Our own observations were complemented with data taken from
the literature (Kennicutt \& Kent 1983, Romanishin 1990, Gavazzi et al. 1991, Young et al.
1996; Koopmann et al. 2001).\\
Furthermore we performed a NIR imaging survey of the same regions (Gavazzi et al. 2000b and
references therein), providing informations on the old stars.\\
In the present paper we combine H$\alpha$ with NIR measurements to study
the young and the old components of the stellar population integrated over the
whole galaxies and we analyze the
properties of the stars as a function of the clustercentric projected distance, of the luminosity and 
of the global gas content.
We postpone to a forthcoming paper the morphological aspects of the analysis related to
the spatial distribution of the young/old stars. 
The present paper is organized as follows: 
in Section 2 we briefly present the new H$\alpha$ imaging observations of 13 galaxies.   
The sample used in the present investigation is illustrated in Section 3.
After defining the "birth-rate" parameter (Sect. 4.1) and the "gas-deficiency" parameter (Sect. 4.2),
we analyze in Section 5.1 the clustercentric dependence of the current star formation rate.
In Sections 5.2 and 5.3 we study
the current star formation properties of galaxies in 3 local clusters
as a function of their global luminosity and gaseous properties.
The conclusions are briefly discussed in Section 6 and summarized in Section 7. 

\section{New observations}

Narrow band imaging in the H$\alpha$ emission line ($\lambda$ = 6562.8 \AA) of 13 galaxies was
obtained in march 20, 2002, using the 2.1m telescope at San Pedro 
Martir Observatory (SPM) (Baja California, Mexico). \\
The target galaxies are listed in Table 3 as follows:
\begin{itemize}
\item{Column 1: VCC (Binggeli et al. 1985) or CGCG (Zwicky et al. 1961-68) designation.}
\item{Column 2: NGC/IC name.}
\item{Column 3: UGC name.}
\item{Columns 4 and 5: J2000 celestial coordinates.}
\item{Column 6: photographic magnitude as given in the VCC or in the CGCG.}
\item{Column 7: heliocentric velocity (km s$^{-1}$) from the VCC or from Gavazzi et al. (1999a).}
\item{Column 8: exposure times in minutes for the ON-band filter.}
\item{Column 9: transmissivity ($\rm R_{ON}$) of the ON-band filter at the redshifted H$\alpha$ line.}
\end{itemize}
We used a Site 1024$\times$1024 pixels CCD detector with pixel size of 0.31 arcsec.
Each galaxy was observed through a narrow band interferometric filter ($\sim 90$ ~\AA ~width) centered at
$\lambda$ 6603, for the galaxies at the redshift of Virgo ($\rm 350<V<3000 ~km~sec^{-1}$), 
and at $\lambda$ 6723 ~\AA, for galaxies in the Coma supercluster.
These observations provided us with the ON-band images and required 15-20 min integration time.
The OFF-band images were obtained through the r-Gunn filter 
and were exposed one fifth of the ON-band ones.
\noindent
The observations were obtained with the seeing ranging from 1.2 to 3 arcsec, but in photometric conditions.
They were flux calibrated using the standard stars Feige 34 and Hz44 from the catalogue of 
Massey et al. (1988), 
observed every 2 hours. Repeated measurements
gave $<$ 0.05 mag differences, which we assume as the typical uncertainty ($1 \sigma$) of the 
photometric results given in this work. \\
The reduction of the CCD frames follows a procedure identical to the one 
described in previous papers of this series (e.g. Gavazzi et al. 2002), based on the IRAF
STSDAS \footnote{IRAF is the Image Analysis and
Reduction Facility made available to the astronomical community by the
National Optical Astronomy Observatories, which are operated by AURA,
Inc., under contract with the U.S. National Science Foundation. STSDAS
is distributed by the Space Telescope Science Institute, which is
operated by the Association of Universities for Research in Astronomy
(AURA), Inc., under NASA contract NAS 5--26555.} 
reduction packages, and it will be briefly summarized here. 
To remove the detector response each image was
bias subtracted and divided by the median of several flat field exposures obtained on
empty regions of the twilight sky. 
Cosmic rays were removed either
using the task COSMICRAY in IRAF or manually by direct inspection of the frames.
The sky background was determined in each frame in concentric object-free regions 
around the galaxies and then subtracted from the flat-fielded images. 
The typical uncertainty on the mean background is estimated 
10 \% of the rms in the individual pixels. This represents the dominant source of 
error in low S/N regions.\\
H$\alpha$ Fluxes and equivalent widths are estimated subtracting the contribution of the continuum
from the ON-band measurements.
As the continuum was estimated using the broad band $r$ filter, which in fact includes the 
H$\alpha$ and [NII] lines, the corrected fluxes and equivalent widths are computed
according to eq. (1) and (2) of paper III, and their uncertainties
are given by:

\begin{equation}
{\sigma_{F_c} = \sigma_{F}\bigg(1+{\frac{\int R_{ON}(\lambda)d\lambda}{\int R_{OFF}(\lambda)d\lambda}}\bigg) }
\end{equation}

\begin{displaymath}
{\sigma_{E.W._c} = \sigma_{E.W.} \bigg(1+{\frac{\int R_{ON}(\lambda)d\lambda}{\int R_{OFF}(\lambda)d\lambda}}\bigg)}
\end{displaymath}
\begin{equation}
{\times \bigg(\frac{1}{1 - (\frac {E.W.}{\int R_{OFF}(\lambda)d\lambda})^2}\bigg)}
\end{equation}

Galaxies with substantial $H\alpha+[NII]$ structure are given in Fig.\ref{results}. The contours of the OFF frames are
superposed to the NET (ON-OFF) frames (grey-scale).

\section{The sample}

\begin{figure*}[!t]
\centerline{
\psfig{figure=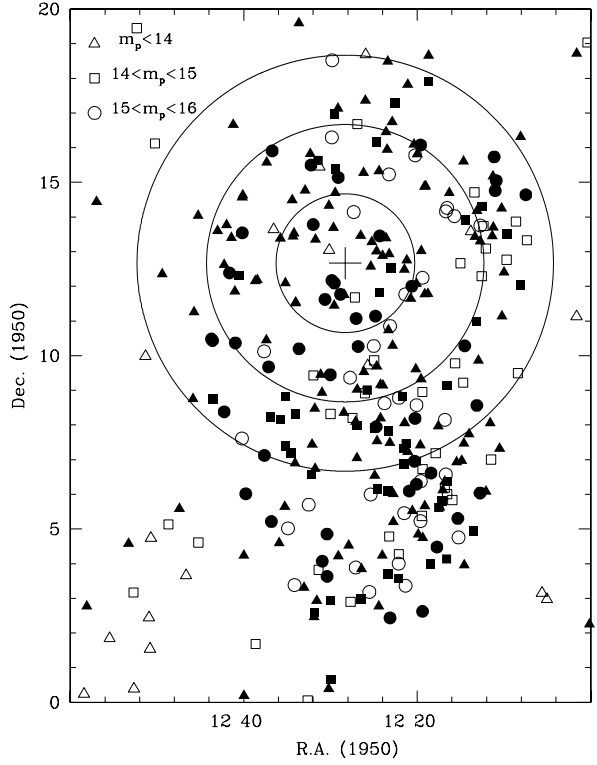,width=11cm,height=11cm}}
\caption{Sky distribution of the 312 spiral galaxies brighter than $m_p \leq 16.0$ in the VCC. 
The filled symbols represent 235 galaxies with 
available H$\alpha$ data, the empty ones to unobserved galaxies. 
Circles are drawn at 2, 4, 6 deg. projected
radial distance from M87 (cross).}\label{vcccelestial}
\end{figure*}

\begin{figure*}[!t]
\centerline{
\psfig{figure=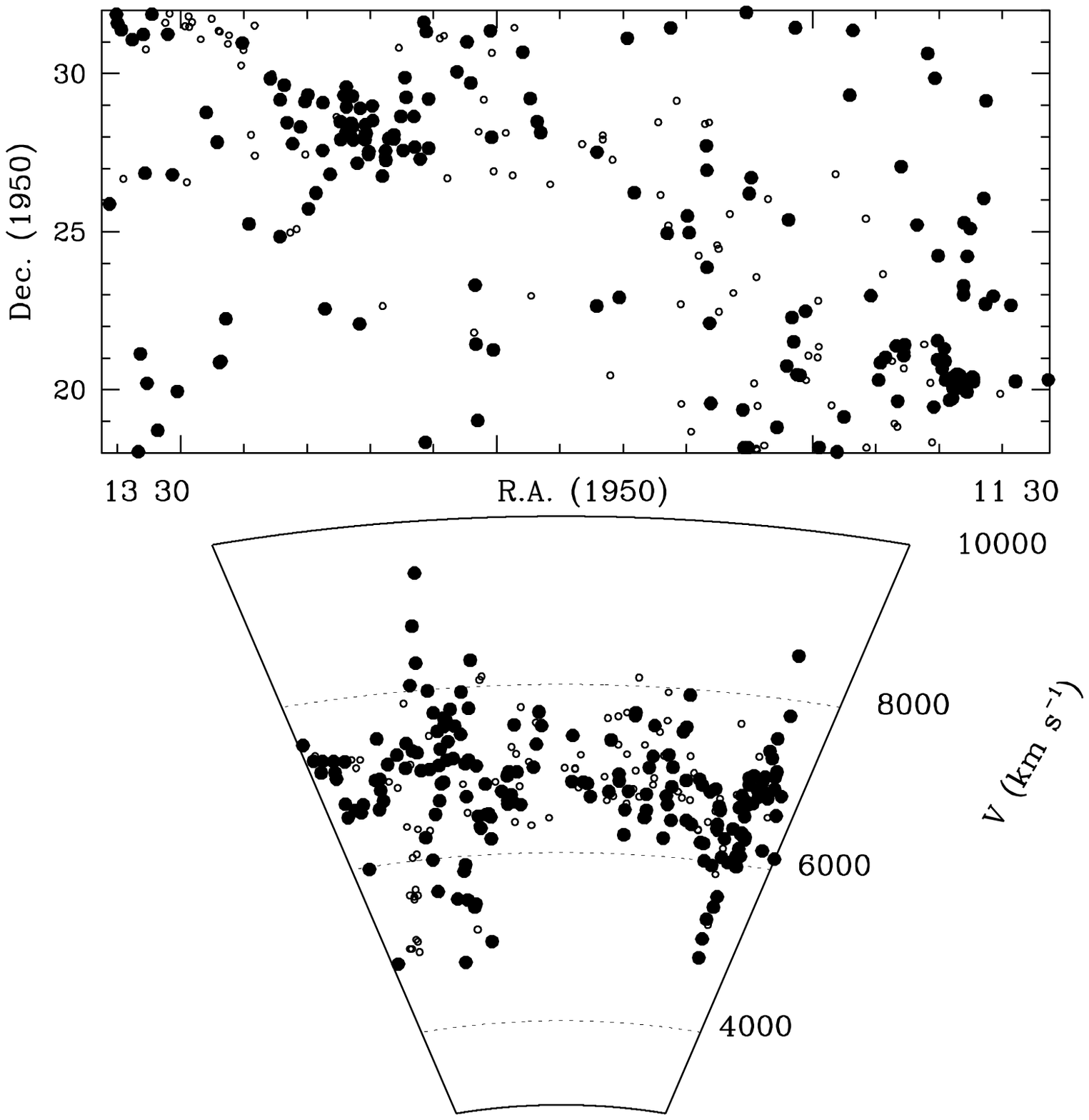,width=11cm,height=11cm}}
\caption{Sky distribution of the 256 spiral galaxies brighter than $m_p \leq 15.7$ in the CGCG
in the Coma supercluster region (top). Wedge diagram (bottom). The filled symbols represent 158
galaxies with available H$\alpha$ data, the empty ones to unobserved galaxies.}\label{comascelestial}
\end{figure*}

\noindent
Including the new observations presented in this paper, this work comprises 
H$\alpha$ and NIR (H band) imaging observations of 369 late-type galaxies belonging to
the Virgo cluster and to the Coma supercluster region.\\
The Virgo cluster galaxies were selected from the Virgo Cluster
Catalogue (VCC) of Binggeli et al. (1985), with $m_p \leq 16.0$,  
Hubble type later than S0a (as given in the VCC) and
classified as cluster members, possible members or belonging to the
W, W', M clouds or to the southern extension (Binggeli et al. 1985; 1993; see also
Gavazzi et al. 1999a) matching $V<3000$  $\rm km ~s^{-1}$ (see Fig. \ref{vcccelestial}).\\
The late-type ($> S0a$) galaxies in the Coma supercluster region
($\rm 18^o \le \delta \le 32^o$; $\rm 11.5^h \le \alpha \le 13.5^h$)
were selected from the CGCG catalogue ($m_p \leq ~15.7$) (Zwicky et al. 1961-68)
and include members to the Coma Supercluster according to Gavazzi et al. (1999b) (see Fig. \ref{comascelestial}).
Table 1 gives the details of the sample completeness in the two studied regions.
The Coma supercluster members are divided in cluster (A1367+A1656) members, members
to groups and pairs (see Gavazzi et al. 1999b) and strictly isolated
supercluster objects (with projected separations $>300$ kpc).
The H$\alpha$ observations were taken either from the present series of papers (Paper I, II, III, IV, primarily
devoted to the Virgo cluster), from Gavazzi et al. (1991, 1998) (containing mostly observations
of the Coma supercluster region) or from 
Kennicutt \& Kent (1983), Kennicutt, Bothun \& Schommer (1984), Romanishin (1990), 
Koopmann et al. (2001) (see detailed references in Table 4).\\
The NIR observations were taken from the series of papers 
"Near-infrared H surface photometry of galaxies" (Gavazzi et al. 1996a, b, 2000a,
Boselli et al. 1997; Boselli et al. 2000 and from Gavazzi et al. 2001). 
Total asymptotic H band magnitudes were obtained by Gavazzi et al. (2000b) and 
by Gavazzi et al. (2001). \\
As listed in Table 1 the combined NIR+ H$\alpha$ observations cover more than
60 \% of the targets in all regions (except Coma supercluster groups+pairs), thus our data can
be considered as representative of the late-type galaxies in the studied regions.\\

\begin{table*}
\caption{The sample completeness}
\label{Tab1}
\[
\begin{array}{lcccc}
\hline
\noalign{\smallskip}
{\rm region} & {m_p \leq 16.0} &  {\rm with~NIR} &  {\rm with~NIR~\&~H_\alpha} & {\rm Compl.}\\
\noalign{\smallskip}
\hline
\noalign{\smallskip}
\rm Virgo               & 323 & 271   &  205 & 63 \%\\
\rm Coma S.~(Clusters)  & 72  & 72    &  54  & 75 \%\\
\rm Coma S.~(Grps+Prs)  & 67  & 67    & 27   & 40 \%\\
\rm Coma S.~(Isolated)  & 119 & 83    & 83   & 69 \%\\
\rm Tot.                & 568 & 480   & 356  & 63 \%\\ 

\noalign{\smallskip}
\hline
\end{array}
\]
\end{table*}

The analyzed galaxies are listed in Table 4 as follows:\footnote{Table 4 is only available in electronic form at http://www.edpsciences.org}

\begin{itemize}
\item{Column 1: VCC designation, from Binggeli et al. (1985) for Virgo galaxies,
or CGCG (Zwicky et al. 1961-68) for Coma supercluster galaxies.}
\item{Column 2: the membership to a cluster or supercluster, 
defined as in Gavazzi et al. (1999a) for Virgo and Gavazzi
et al. (1999b) for the Coma/A1367 supercluster}
\item{Column 3: morphological type as given in the VCC or in Gavazzi \& Boselli (1996).}
\item{Column 4: projected angular separation from the nearest cluster center (degrees).}
\item{Column 5: asymptotic H band magnitude, obtained
as described in Gavazzi et al. (2000b).}
\item{Column 6: distance (Mpc); we assume 17 Mpc for Virgo A, N, S, E, 23 Mpc for
Virgo B and 32 Mpc for Virgo W, M as given in Gavazzi et al. (1999a), 96 Mpc for Coma, 
91 Mpc for A1367. For galaxies not belonging to the clusters, the distance is determined
from the redshift using $H_0$= 75 $\rm km~s^{-1}Mpc^{-1}$.}
\item{Column 7: gas deficiency parameter as defined in Section 4.2.}
\item{Column 8: $H_\alpha+[NII] E.W.(\AA)$.}
\item{Column 9: Log of the $H_\alpha$ flux ($\rm erg ~cm^{-2} ~s^{-1}$) deblended from the [NII]
contribution and corrected for internal extinction as in Boselli et al. (2001).}
\item{Column 10: reference to the $H_\alpha$ data (as listed at the bottom of the Table).}
\end{itemize}


\section{Tools}

\subsection{The birthrate parameter}

\begin{figure}[]
\psfig{figure=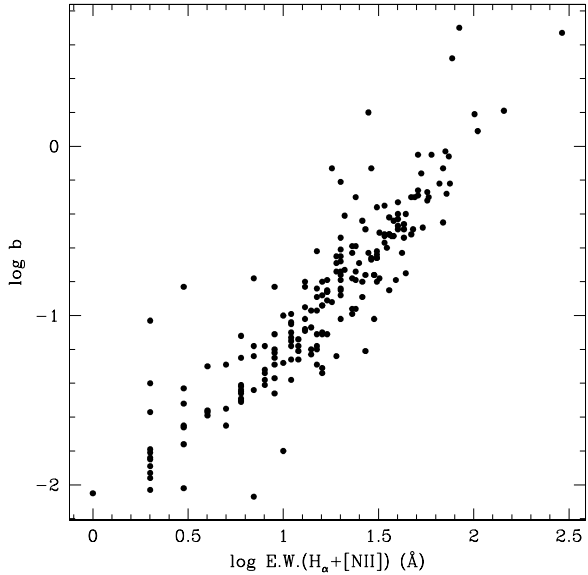,width=9cm,height=9cm}
\caption{The relation between the birthrate parameter and the $H_\alpha$ emission
line equivalent width.}\label{bEW}
\end{figure}

$H_\alpha$ and NIR observations provide us with information on 
stellar populations with different time scales: $\sim 10^7$ yrs the former
and $\sim 10^{10}$ yrs the latter. The two quantities combined give the ratio 
of the current SFR to the average past SFR or  
the birthrate parameter $b$, as defined by Kennicutt et al. (1994).\\
Following Boselli et al. (2001), we use the Near Infrared luminosity $L_H$ as 
a tracer of the global mass of old stars, assuming that disk galaxies have a constant $M_{Tot}/L_H=4.6$
within their optical radius 
(Gavazzi et al. 1996c). Thus we write the adimensional parameter $b$ as:
\begin{equation}
b_{obs} = \frac{SFR~t_o~(1-R)}{L_H ~(M_{Tot}/L_H)~DM_{cont}} 
\end{equation}
where SFR is derived from the $H_\alpha$ luminosity with:
\begin{equation}
{SFR [M\odot yr^{-1}] = K_{H\alpha} L_{H\alpha} [erg~s^{-1}]}
\end {equation}
Obviously the $H_\alpha$ luminosity is deblended from the observed [NII]
contribution and corrected for internal extinction as in
Boselli et al. (2001).  For consistency with Boselli et al. (2001) we
adopt $K_{H\alpha}= 1/1.16~10^{41}$ for an IMF of slope -2.5 in the mass range
0.1--80 $M\odot$.\\
$DM_{cont}$ is the dark matter contribution at the optical radius,  i.e. within the
$25 ~mag~ arcsec^{-2}$ B band isophote, that we assume 
$DM_{cont}$=0.5, as in Kennicutt et al. (1994).\\
$R=0.3$ (Kennicutt et al. 1994) is the fraction of gas that stars re-injected 
through stellar winds into the interstellar medium during their lifetime,
that we assume $t_o$ $\sim$12 Gyrs.\\
If we assume that galaxies evolved as "closed" systems following 
an exponential  
Star Formation History (SFH), with a characteristic decay time $\tau$
since their epoch of formation ($t_o$), their birthrate parameter
can be computed analytically (see Boselli et al. 2001) as:
\begin{equation}
b_{mod} = \frac{t_o ~e^{-t_o/\tau}}{\tau (1-~e^{-t_o/\tau})}
\end{equation}
$b_{mod}$ can be written as a function of $L_H$ 
using the relation between $\tau$ and $L_H$ found by Boselli et al. (2001): 
\begin{equation}
log \tau = -0.4 (Log L_H - 12 ) [\rm Gyr]
\end{equation}
where
\begin{equation}
Log L_H = 11.36 -0.4 H + 2 log (Dist) [\rm L_H \odot]
\end{equation}
The dependence of $b_{mod}$ on $L_H$ is given as 
a dotted line in Figs.\ref{comas}, \ref{virgo_def} and \ref{nondefgas}.\\
Although $b$ and $H_\alpha$ E.W. have distinct dimensions, they are strongly correlated quantities.
In fact they are operationally obtained in a similar way: $b$ is computed by normalizing the $H_\alpha$ line intensity
to the NIR continuum intensity, while the equivalent width is divided by the continuum intensity underlying the
$H_\alpha$ line. This is shown in Fig.\ref{bEW} which can be
directly compared with Fig.4 of Kennicutt et al. (1994). 
 
\subsection{The global gas deficiency parameter}

\begin{figure}[]
\psfig{figure=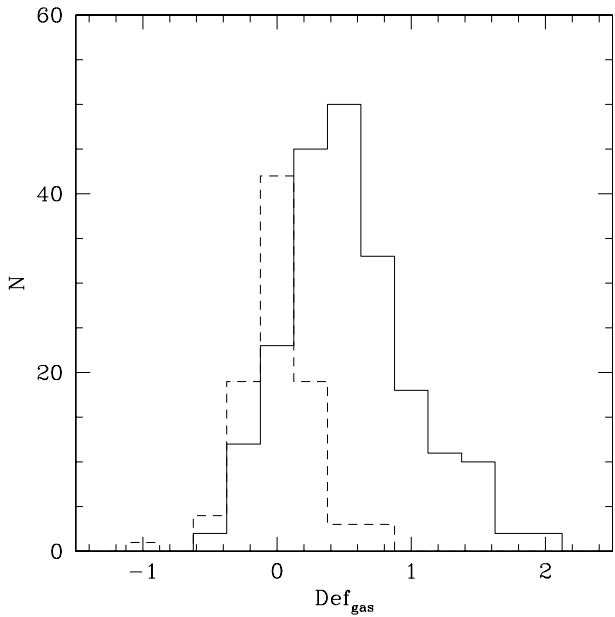,width=9cm,height=9cm}
\caption{Histograms of the $Def_{gas}$ parameter for the isolated galaxies 
in the Coma supercluster (dashed line) and 
for Virgo galaxies (continuous line).}\label{histogas}
\end{figure}

For galaxies in our sample we estimate the "global gas content"
$M_{gas}=M_{HI}+M_{H2}+M_{He}$.\\
$M_{HI}$ is available for most (95 \%) targets by direct 21 cm observations
(see Scodeggio \& Gavazzi 1993, Hoffman et al.
1996, and references therein).
The mass of molecular hydrogen can be estimated from the measurement 
of the CO (1-0) line emission, assuming a conversion factor ($X$) 
between this quantity and the $H_{2}$ surface density. 
$X$ is known to vary in the
range $10^{20}$ to $10^{21}$  [mol cm$^{-2}$ (K km s$^{-1}$)$^{-1}$]
from galaxy to galaxy, according to their metallicity and UV radiation field. We adopt the empirical calibration
as a function of the H band luminosity:
\begin{equation}
log X = 24.23 - 0.38*log L_H 
\end{equation}
found by Boselli et al. (2002a).
The CO (1-0) line emission is unfortunately available for 52 \% of the 
considered sample (see Boselli et al. 2002a and references therein), 
and it is assumed 15 \% of the HI content for the remaining
objects (as concluded by Boselli et al. 2002a).\\
The contribution of He, not directly observable, is estimated as
30 \% of $M_{HI}+M_{H2}$ (see Boselli et al. 2002a).\\
We define the "gas deficiency" parameter $Def_{gas}= Log M_{gas~ref.} - Log M_{gas~obs.}$
as the logarithmic difference between $M_{gas}$ of a reference sample of isolated 
galaxies and $M_{gas}$ actually observed in individual objects (in full analogy with 
the definition of HI deficiency by Giovanelli \& Haynes 1985).
Using a procedure similar to the one adopted by Haynes and Giovanelli (1984)
we find that the gas content of 72 isolated objects
in the Coma Supercluster correlates with
their linear optical diameter (D):
$Log M_{gas~ref}=a+b Log(D)$,
where $a$ and $b$ are weak functions of the Hubble type, as listed in Table 2. 
$Def_{gas}$ are listed in Column 7 of Table 4. 
Histograms of the $Def_{gas}$ parameter are given in Fig. \ref{histogas} for the Coma isolated 
objects and for the Virgo galaxies. Isolated objects have $Def_{gas}=0 \pm 0.18$, 
while Virgo galaxies have significantly positive $Def_{gas}=0.53 \pm 0.35$.

\section{Results}

\begin{figure}[t]
\psfig{figure=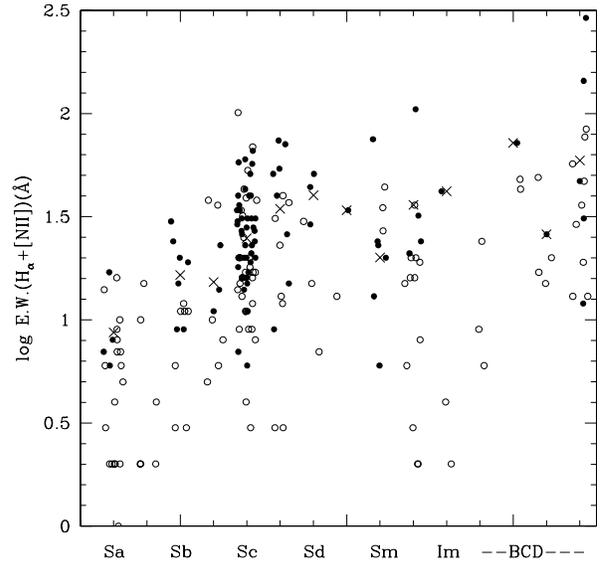,width=9cm,height=9cm}
\caption{The distribution of $H_\alpha$ E.W. of spiral galaxies in the Virgo cluster 
as a function of Hubble type. Filled dots represent galaxies with normal gas content ($Def_{gas}<0.4$),
open symbols are gas deficient objects. To avoid superposition of points, galaxies in each type bin 
are separated by a small random quantity.
Crosses represent averages (including only the non gas-deficient galaxies) in each morphological type bin.}\label{type}
\end{figure}

The $H_\alpha$ E.W. of galaxies is known to increase systematically along the Hubble sequence,
from virtually zero for the early types (E-S0) to several hundred \AA~ 
for the latest types (Kennicutt 1998). A weak trend is confirmed when data 
limited to the Virgo spiral galaxies included in this work are used, as shown in Fig.\ref{type}.
However the scatter in each of the morphological type bins is as much as an order of magnitude,
even though the scatter is somewhat reduced when gas deficient galaxies are excluded.
The Hubble type alone does not account for the star formation properties of galaxies in this cluster.
To shed light on other possible dependences we will analyze how the SFR varies as a function of
the projected clustercentric distance (sect. 5.1), of the luminosity (sect. 5.2) 
and of the gaseous content of galaxies (sect. 5.3).      

\subsection{The clustercentric dependence of the SFR}

\begin{figure}[t]
\psfig{figure=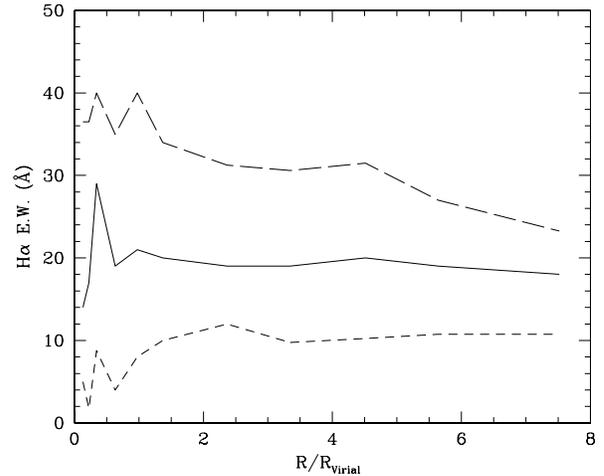,width=9cm,height=9cm}
\caption{The distribution of $H_\alpha$ E.W. as a function of (projected) clustercentric radius from the Coma
and A1367 clusters ($M_b<-19$). The top and bottom lines represent the $75^{th}$ and the $25^{th}$ percentile
of the EW distribution, while the central line is the median of the distribution.}\label{comas_virial}
\end{figure}

\begin{figure}[t]
\psfig{figure=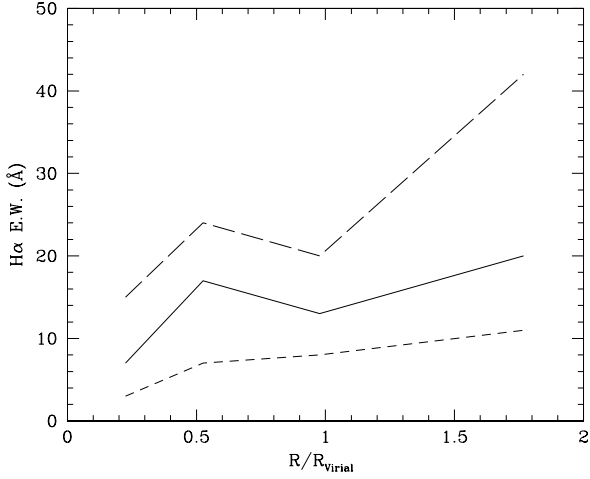,width=9cm,height=9cm}
\psfig{figure=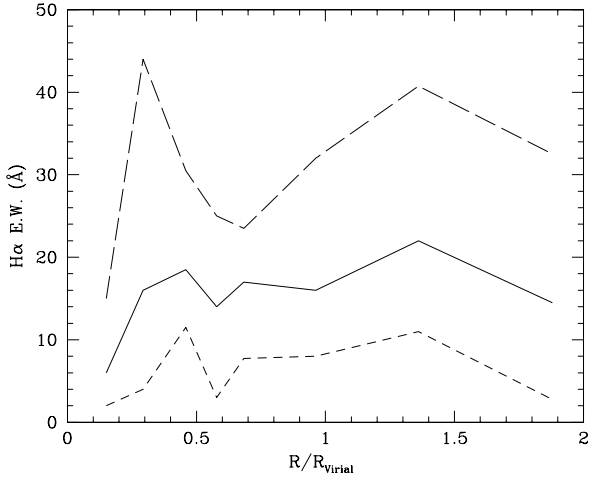,width=9cm,height=9cm}
\caption{The distribution of $H_\alpha$ E.W. as a function of (projected) clustercentric radius from 
the Virgo cluster. The top and bottom lines represent the $75^{th}$ and the $25^{th}$ percentile
of the EW distribution, while the central line is the median of the distribution. The top panel shows
the Virgo galaxies brighter than $M_p=-19$, while the bottom panel includes all galaxies surveyed in $H_\alpha$
($M_b<-15$).}\label{virgo_virial}
\end{figure}

Lewis et al. (2002) analyzed the dependence of the galaxy SFR on the 
projected distance from clusters in the 2dF survey.
Their volume limited samples comprise galaxies of all morphological types with $0.05<z<0.1$, 
brighter than $M_b<-19$. 
They showed with high statistical significance that the median SFR
of galaxies decreases with decreasing projected distance from clusters.\\
It would be interesting to compare these intermediate distance clusters 
with the 3 local clusters analyzed in this work, however a direct comparison
cannot be carried out because data of early-type galaxies are not in our possession. 
The dependence of the $H_\alpha$ E.W. on the clustercentric distance in units of virial radii
can be analyzed only for the late-types galaxies, bearing in mind that our completeness is 60 \%.
We compute $R_{virial}=0.002\sigma_r h^{-1}$ (Girardi et al. 1998)
for the 3 clusters assuming $\sigma_r$ = 775, 840, 925 $\rm km~sec^{-1}$
for Virgo, A1367 and Coma respectively.\\
The combined Coma and A1367 clusters (with $M_b<-19$) are shown in Fig. \ref{comas_virial} enbedded in the Coma supercluster 
that we trace out to large clustercentric radial distances.
We find a significant inner decrease only of the $25^{th}$ percentile of the $H_\alpha$ E.W.
distribution. Both the median and the $75^{th}$ percentile instead increase inwards.
We find it unlikely that the $H_\alpha$ E.W. distribution is biased toward high values due to incompleteness,
since for the Coma+A1367 clusters our survey covers 75 \% of the sample.
These clusters are inhabited by strong $H_\alpha$ emitters
to which the attention has been drawn by several authors. These include the "blue galaxies in the Coma cluster"
of Bothun \& Dressler (1986) and the blue galaxy sample observed with ISO by Contursi et al. (2001).
Many (13) galaxies with $H_\alpha$ E.W. in excess of 50 \AA~ are found both in the inner regions ($R/R_{virial}<0.5$) 
and at intermediate distances ($0.5<R/R_{virial}<1.5$) from the observed clusters.
Noticeably these galaxies are near the faint limit of our survey ($-19.5<M_b<-19$ mag).\\
For the Virgo cluster we separate the bright sample
($M_p<-19$), with a luminosity cutoff and $H_\alpha$ completeness similar to the Coma supercluster (75 \%), 
from the total sample
($M_p<-15$) and we show the two radial dependences separately in Fig.\ref{virgo_virial}.
The bright sample shows an inner decrease of the SFR. For the total sample this
pattern no longer holds true. The Virgo cluster contains 24 galaxies with $H_\alpha$ E.W. in excess of 50 \AA~ 
(11 are BCDs), the majority (14/24 objects) being fainter than -17.1 mag.\\ Because of their low optical luminosity
the strong $H_\alpha$ emitters belonging to Virgo would have all escaped detection in 
the 2dF survey.
We conclude that, beside morphology segregation, the three local clusters analyzed in this work 
do not show a clear radial trend of the SFR distribution. The presence of the radial trend depends purely 
on a luminosity cutoff, which varies cluster to cluster between -17 and -19 mag.  
While spiral galaxies brighter than this cutoff luminosity have lower than average SFR at 
the cluster centers, galaxies fainter than this limit have SFR independent from the clustercentric projected distances. 
This is consistent with the idea that infall of small galaxies is occurring onto 
rich clusters at the present cosmological epoch.

\subsection{The SFR in the Coma supercluster}

\begin{figure}[t]
\psfig{figure=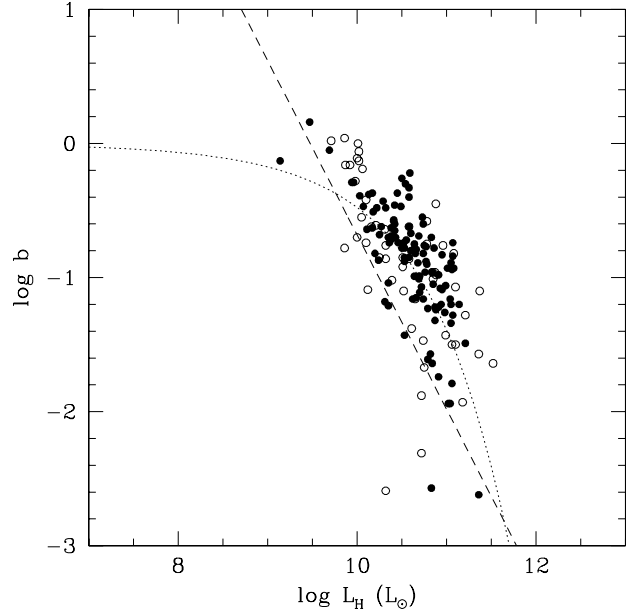,width=9cm,height=9cm}
\caption{The relation between the birthrate parameter and the NIR 
luminosity (mass) for the Coma supercluster galaxies. Galaxies in 
the Coma+A1367 clusters are represented with empty symbols, filled 
symbols are non-cluster galaxies.
The dotted line represents the expected $b$ as a function of $L_H$ in the
closed-box model of equation (5). The dashed line represents the observational bias 
affecting the Coma galaxies due to their selection in the B band.}\label{comas}
\end{figure}
Since, as concluded in the previous section, the present star formation rate of galaxies
near the center of the studied clusters is a luminosity sensitive parameter, it is compelling
to proceed to a systematic investigation of the luminosity dependence of the star formation properties.
To this aim it is adequate to analyze the luminosity dependence of the birthrate parameter $b$ (see section 4.1).
The most appropriate luminosity indicator, which we will adopt hereafter, is the NIR (H band) luminosity. This  
parameter traces at best the dynamical mass (within the optical disk) of spiral galaxies,
as concluded by Gavazzi et al. (1996c), who found $Log M_{dyn}= Log L_H + 0.66$.\\
The dependence of the $b$ parameter on $L_H$, given in Fig. \ref{comas}, 
shows that the star formation history of spiral galaxies in the 
Coma supercluster region is in almost inverse proportionality with 
the system luminosity (mass). The most massive spirals ($\rm Log L_H \sim 11.5 L_H \odot \sim 12.3 M\odot$ ) 
have their $b$ parameter as much as 100 times lower than less luminous 
(giant) galaxies ($\rm Log L_H \sim 10 L_H \odot \sim 10.8 M\odot$). This confirms previous claims that the 
current SFR, as derived from the $H_\alpha$ E.W., anti-correlates with 
the system mass (Gavazzi et al. 1998).
Furthermore Fig.\ref{comas} shows that there is not a significant 
difference between the SFH of galaxies in the rich Coma+A1367 
clusters and of relatively isolated objects in the same supercluster, 
in agreement with Gavazzi et al. (1998).\\ 
Both results are however biased by selection effects.
The Coma supercluster galaxies were selected optically in the blue 
(photographic) band ($m_p \leq ~15.7$). The selected targets were observed "a posteriori" 
in $H_\alpha$ and in the NIR, therefore at any given $L_H$ only 
galaxies bluer than a certain threshold are sampled. In other words 
the B selection biases against faint-red galaxies, according to the relation between B-H
and the infrared luminosity represented by equation \ref{bhlh} (see Scodeggio et al. 2002). 
This, combined with the fact that
$b$ correlates with the B-H color (see equation \ref{bbh}), introduces a selection effect 
in the $b$ vs. $L_H$ plane 
(see equation \ref{blh}).
These empirically determined relations are:   
\begin{equation}
{B_{lim} - H = -12.7 - 5~log(dist) + 2.5~log L_H} \label{bhlh}
\end{equation} 
where $B_{lim}=-19.2$ corresponds to the limiting magnitude ($m_p \leq ~15.7$) at the Coma distance that 
we assume 96 Mpc. 
\begin{equation}
{log b = 0.56 - 0.52 (B-H)}  \label{bbh}
\end{equation}
\begin{equation}
{log b = 7.16 + 2.6~log(dist) - 1.3~log L_H} \label{blh}
\end{equation}  
Equation \ref{blh} is represented in Fig. \ref{comas} with a dashed line. In conclusion, faint-low star 
forming galaxies at the distance of Coma 
below the diagonal line of Fig. \ref{comas} are severely undersampled.
\begin{table}[!b]
\caption{The $M_{gas}$ vs. diameter relation for isolated galaxies}
\label{Tab 2}
\[
\begin{array}{lccc}
\hline
\noalign{\smallskip}
{Type} & {a} & {b}  & {R^2}\\
\noalign{\smallskip}
\hline
\noalign{\smallskip}
Sa-Sb & 7.62 & 1.55 & 0.75\\
Sbc-Sc & 7.48 & 1.68 & 0.75\\
Scd-Irr & 7.74 & 1.49 & 0.77\\
\noalign{\smallskip}
\hline
\end{array}
\]
\end{table}
\subsection{The SFR in the Virgo cluster}


\begin{figure}[!t]
\psfig{figure=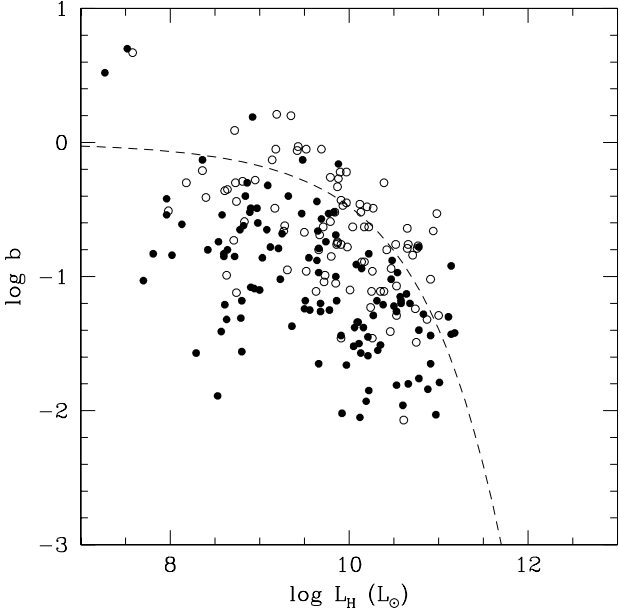,width=9cm,height=9cm}
\caption{The relation between the birthrate parameter and the NIR luminosity (mass)
for the Virgo galaxies. Empty symbols represent galaxies with normal gas content ($Def_{gas}<0.4$)
while deficient galaxies ($Def_{gas}>0.4$) are given with filled symbols.}\label{virgo_def}
\end{figure}

\begin{figure}[!t]
\psfig{figure=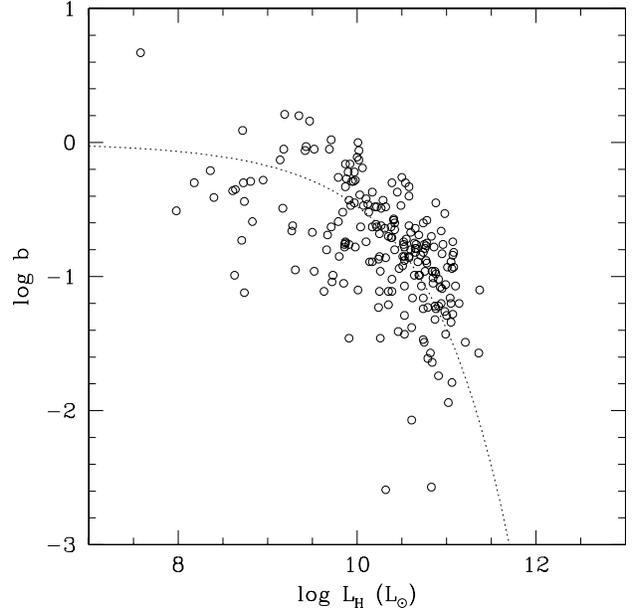,width=9cm,height=9cm}
\caption{The relation between the birthrate parameter and the NIR luminosity (mass)
for Virgo+Coma galaxies with normal gas content ($Def_{gas}<0.4$).}\label{nondefgas}
\end{figure}

\begin{figure}[!t]
\psfig{figure=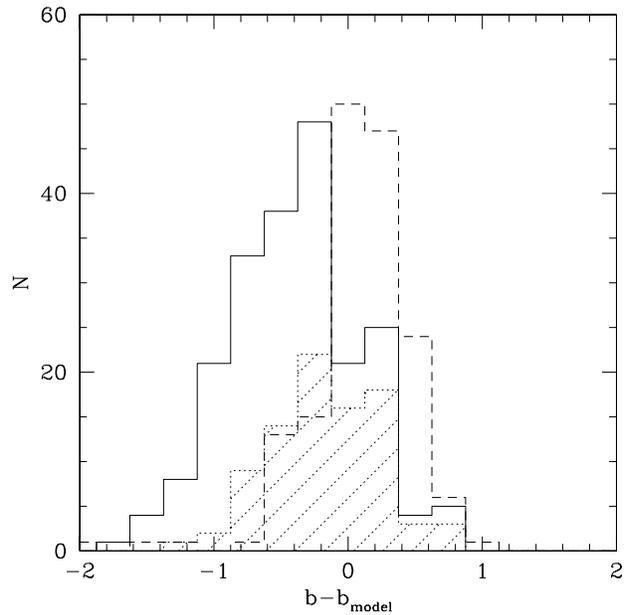,width=9cm,height=9cm}
\caption{Histograms of the residual $b_{obs}-b_{mod}$ for the Coma galaxies (dashed line),
for Virgo (continuous line) and for the Virgo galaxies with normal HI content ($Def_{gas}<0.4$)
(dashed histogram).}\label{histo}
\end{figure}

\begin{figure}[!t]
\psfig{figure=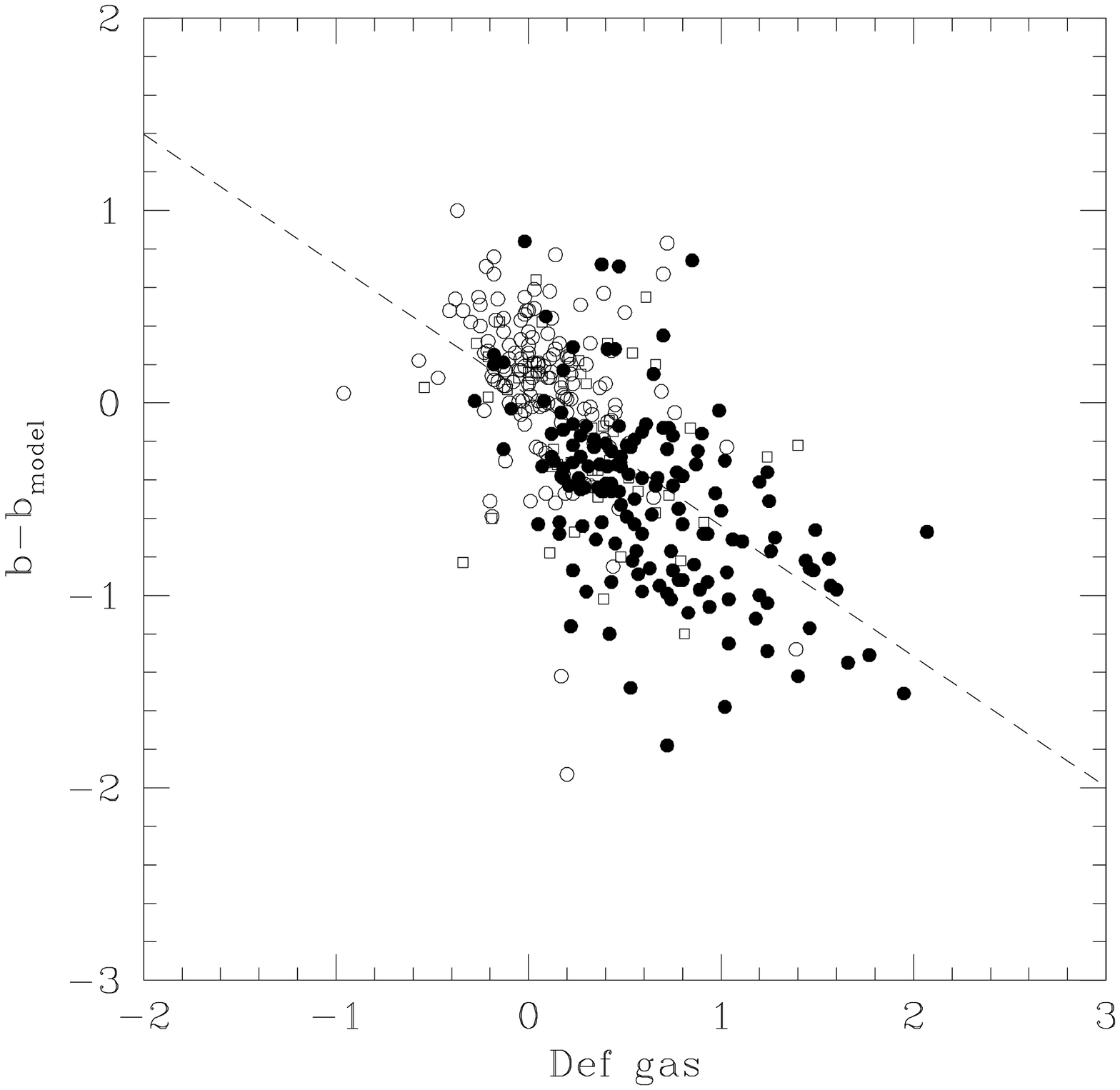,width=9cm,height=9cm}
\caption{The relation between $b_{obs}-b_{mod}$ and the $Def_{gas}$ parameter. Empty circles
are Coma supercluster galaxies, empty squares are "normal" Virgo clouds (N, W, S), filled circles
are "deficient" Virgo clouds (A, B, E).}\label{resid}
\end{figure}

The selection effect mentioned above affects the Virgo sample 
to a much lesser extent, because Virgo is 3.7 magnitudes closer than Coma. 
When we consider the Virgo galaxies alone in Fig. \ref{virgo_def} 
we include dwarf systems with $L_H$ fainter by almost 2 orders of magnitudes with respect to Coma. 
The scatter of the $b$ vs. $L_H$ relation increases considerably because 
the large majority of faint Virgo objects have $b$ lower than Coma. This is 
in agreement with Kennicutt (1983) who found evidence for significant $H_\alpha$
deficiency in 12 Virgo galaxies with respect to isolated galaxies. 
Galaxies with $b$ as low as the ones in Virgo might exist in the Coma+A1367 clusters as well, 
but are not observed because of the previously mentioned observational bias. 
Thus we conclude that, at any given mass, spirals belonging to the Virgo cluster 
have their present star formation activity significantly lower than  
isolated galaxies.\\
It remains to be explained why. The first thing to explore is whether 
their gaseous content is sufficient for fueling the star formation. 
Cluster spirals are in fact known to suffer from HI deficiency 
(Giovanelli \& Haynes 1985; Solanes et al. 2001), a pattern
that is interpreted in the framework of the ram pressure mechanism 
(Gunn \& Gott, 1972).\\
When galaxies are separated according to their 
gas deficiency parameter (see Fig. \ref{virgo_def}), we recognize that, at any given $L_H$, 
galaxies with "normal" gas content ($Def_{gas}<0.4$)(open symbols) 
have their $b$ parameter significantly higher than gas "deficient" objects.\\
Fig. \ref{nondefgas} is restricted to the non deficient galaxies of both the Virgo and Coma regions. 
In this and in the previous figures the dotted curve represents $b_{mod}$ i.e. 
the $b$ vs. $L_H$ relation expected from the closed-box scenario, in the 
assumption that $\tau$ is inversely proportional to $L_H$ according to equation (6). 
Galaxies in Fig. \ref{nondefgas} are found in relatively good agreement with $b_{mod}$, 
in other words their residuals $b_{obs} - b_{mod}$ are small.
This is evidenced in the histograms of Fig. \ref{histo} where the distribution 
of the residuals $b_{resid}= b_{obs} - b_{mod}$ is given separately for the 
Coma galaxies, for the Virgo galaxies and for the subsample of the Virgo 
galaxies with normal gas content ($Def_{gas}<0.4$). Large negative residuals,
implying a factor of 3 lower SFR, 
are associated with significantly gas deficient galaxies. It is concluded that,
at any given luminosity,  
the principal parameter regulating the current star formation activity in cluster spirals is the 
availability of gas at their interior. 
This is further evidenced in Fig. \ref{resid}, where 
$b_{resid}$ is plotted against the gas deficiency parameter, 
showing a significant linear anti-correlation: $b_{resid}=0.04-0.68 \times Def_{gas}$.

\section{Discussion and conclusions}

\begin{figure}[!t]
\psfig{figure=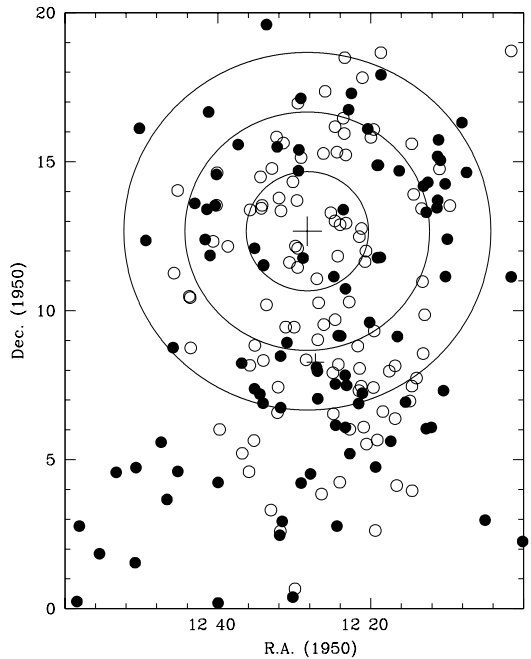,width=10cm,height=10cm}
\caption{The distribution of the "quenched" (empty symbols) and "healthy" (filled symbols) galaxies
in the Virgo cluster. Positions of M87 and M49 are shown by crosses.}\label{vccdef}
\end{figure}

We have shown that a large fraction of late-type galaxies in the Virgo cluster
have their current star formation rate significantly quenched with respect to
isolated objects. These systems coincide with the Virgo gas deficient galaxies.
Since the "gas" deficiency parameter is dominated by 
the HI phase (H$_2$ contributes only to 15 \% of the HI), it is concluded that, to the first order, 
the star formation properties of galaxies in the Virgo cluster are determined by 
the pattern of HI deficiency.
As earlier recognized by Kennicutt (1998), this is a somewhat surprising result, 
because the typical scales of HI and of star formation are
very different in disk galaxies. HI reservoirs extend some 2 x the scale where the star formation takes place 
(Cayatte et al. 1994). We will re-examine this issue in more details in our forthcoming
paper dedicated to the morphology of the star formation regions in galaxies,
where a comparison between the scale-length of $H_\alpha$, $HI$ and $H_2$ will be
carried on specifically.   
Limiting ourself to the global quantities, they indicate that
infall of HI gas occurs in the disks on time scales similar to the star formation time. 
If the gas replenishment fails, because the HI reservoir is reduced by some 
ablation mechanism (e.g. ram pressure), the star formation
adjusts itself to significantly lower rates.\\
Galaxies with  $b_{resid}< -0.7 $ and $Def_{gas}>0.4$ ("quenched") are plotted in Fig.\ref{vccdef} with 
empty symbols, together with their "healthy" counterparts (filled symbols).
Beside a marginal clustering of deficient objects around M87 (cluster A) and M49 (cluster B) 
the two populations appear mixed in position. There is for example a considerable fraction
of "healthy" objects projected onto the center of cluster A.
However Virgo is known to be a complex dynamical entity, composed by the main cluster (A)
a secondary cluster (B), several Mpc behind A, and a number of clouds at approximately the
distance of A, but with significantly discrepant velocities, suggesting infall (see Gavazzi et al. 1999a).\\
By considering galaxies with projected angular separation $< 3.7$ deg from M87
we isolate 68 bona fide members of cluster A. We divide them into 48 "quenched" and 20 "healthy".
For a considerable fraction (22/48 and 13/20 respectively) their distance is available from Gavazzi et al. (1999a)
based on the H band Tully-Fisher relation (Tully \& Fisher 1977) (distances of few galaxies whose H magnitudes were not yet available to
Gavazzi et al. 1999a were recomputed by us). 
To our surprise we find that, while the average distance modulus of the deficient objects ($\mu_o=$30.85)
is in perfect agreement with the  distance modulus of cluster A as a whole ($\mu_o=$30.82) 
(Gavazzi et al. 1999a), the distance modulus of the non-deficient 
galaxies projected onto A is $\mu_o=$31.77 on average, thus
almost one mag more distant. 
It is thus concluded that "healthy" spirals projected onto the
Virgo center belong in fact to a background cloud with a distance comparable with that of cluster B. 
This cloud has not yet entered the dense environment of cluster A.

\section{Summary}
\begin{itemize}
\item{We have covered with H$\alpha$+[NII] and NIR imaging observations $\sim 60 \%$
of the late-type (spiral) galaxies brighter than $m_p=16.0$ 
in the Virgo cluster and in the Coma/A1367 supercluster.}
\item{The H$\alpha$ E.W. of spiral galaxies shows the expected decrease toward the center of the three
studied clusters only when galaxies brighter than -19.5 are considered. Weaker and dwarf systems
show no or reverse trends, indicating that substantial infall of small spiral galaxies is currently
taking place onto local clusters.}
\item{From the combined H$\alpha$ and NIR data we derive the birthrate, i.e. the fraction of young to old stars, providing an
estimate of the star formation history for these galaxies.}
\item{The birthrate parameter shows a weak increasing trend with increasing lateness in the Hubble classification.}
\item{The birthrate parameter of isolated galaxies in the Great Wall is in almost inverse proportionality with the NIR
luminosity, i.e. with the systemic mass. Giant spiral galaxies have a ratio of young-to-old stars 
100 times lower than their dwarf counterparts.}
\item{A large fraction of spiral galaxies in the Virgo cluster have a birthrate parameter significantly
lower (a factor 3) than isolated galaxies of similar luminosity.}
\item{Galaxies with quenched current star formation coincide with galaxies with significant gas deficiency.}
\item{A population of currently star forming galaxies with normal gas content is found projected near
the center of the Virgo cluster. Their Tully-Fisher distance is approximately 1 mag larger than the
one of the deficient objects, which corresponds with the distance of the M87 cluster. This points out
the existence of a distinct cloud of galaxies falling onto the Virgo cluster.}    
\end{itemize}

\begin{acknowledgements}

This research has made use of the NASA/IPAC Extragalactic Database (NED) which is operated 
by the Jet Propulsion Laboratory, California Institute of Technology, under contract with the
National Aeronautics and Space Administration. 
\end{acknowledgements}

\onecolumn
\begin{longtable}{rccrrcrcc}
\caption{The newly observed galaxies}\\
\hline
\noalign{\smallskip}

VCC/CGCG & NGC/IC & UGC &$RA~~(J2000)$&$Dec$  &  $m_{pg}$  & $Vel$  &$T_{on}$&$R_{ON}$  \\
(1) & (2) & (3) & (4) & (5) & (6) & (7) & (8) & (9) \\
\noalign{\smallskip}
\hline
\noalign{\smallskip}
343     &  3148   &   -    & 121921.68  &  075213.8    &    15.1  &  2479   & 20 & 0.79 \\
841     &   -     &   -    & 122547.40  &  145711.4    &    15.6  &  501    & 20 & 0.66 \\
15031   &  4771   &   8020 & 125321.85  &  011613.5    &    13.3  &  1119   & 15 & 0.85 \\
15049   &  4845   &   8078 & 125801.33  &  013430.3    &    12.9  &  1097   & 15 & 0.85 \\
15055   &  4904   &   8121 & 130058.89  & -000142.4    &    13.2  &  1174   & 15 & 0.85 \\
41041   &  4116   &   7111 & 120736.33  &  024133.1    &    13.0  &  1309   & 15 & 0.85 \\
43028   &  4688   &   7961 & 124746.67  &  042005.3    &    14.5  &  984    & 15 & 0.84 \\
43034   &  4701   &   7975 & 124911.87  &  032324.5    &    13.1  &  727    & 15 & 0.79 \\
43054   &  4765   &   8018 & 125314.52  &  042749.4    &    13.0  &  725    & 15 & 0.79 \\
69036   &  4067   &   7048 & 120411.46  &  105114.8    &    13.2  &  2424   & 15 & 0.8  \\
100015  &  4758   &   8014 & 125244.16  &  155050.9    &    14.1  &  1240   & 15 & 0.85 \\
157075  &  -      &   -    & 115940.06  &  263248.7    &    15.7  &  6694   & 15 & 0.77 \\
160121  &  -      &   8161 & 130329.11  &  263300.8    &    15.5  &  6676   & 20 & 0.77 \\
\hline
\end{longtable}
\normalsize  
\newpage


\begin{longtable}{ccccrccccc}
\caption{Parameters of the analyzed galaxies}\\
\multicolumn{10}{c}{Virgo}\\
\hline
\noalign{\smallskip}

Gal         & Agg    &  Type        & $\Theta$      & H      &   Dist.    & $Def_{gas}$ & $H_\alpha+[NII] E.W.$ & $Log F(H_\alpha)$  &  ref\\
	    &        &              & Deg.         & mag    &   Mpc      &          & $\AA$           & $\rm erg ~cm^{-2} ~s^{-1}$   & \\
(1)         &   (2)  &  (3)         &    (4)       &  (5)   &   (6)      &  (7)     &  (8)            & (9)              &  (10) \\
\noalign{\smallskip}
\hline                                                                                                                              
\noalign{\smallskip}

VCC0001  &  $V_M$   &   BCD    &     5.63      &   12.75   &   32   &  -	 &    12   & -13.46   &      11 \\     
VCC0010  &  $V_M$   &   BCD    &     5.35      &   12.76   &   32   &  0.34  	 &    31   & -12.98   &       6 \\     
VCC0017  &  $V_M$   &   Im     &     5.43      &  14.12    &   32   &  0.26  	 &   105   & -12.78   &       9 \\
\hline
\end{longtable}
  \footnotesize{References:}
  \noindent
  (1) Kennicutt \& Kent, 1983; (2) Kennicutt, Bothun \& Schommer, 1984; (3) Gavazzi, Boselli \& Kennicutt, 1991;
  (4) Romanishin, 1990; (5) Gavazzi et al., 1998; (6) Almoznino \& Brosch, 1998; (7) Boselli \& Gavazzi, 2002 (8) Moss, Whittle \& Pesce, 1998;
  (9) Heller, Almoznino \& Brosch, 1999; (10) Usui, Saito, \& Tomita A., 1998; (11) Gavazzi et al. 2002a; (12) Boselli et al., 2002b; 
  (13) Koopmann, Kenney \& Joung, 2001; (14) Iglesias et al., 2002; (T.W.) This work.

\normalsize

\newpage
\setcounter{figure}{13}
\begin{figure}[!b]
\caption{Newly observed galaxies with substantial $H\alpha+[NII]$ structure. The NET
(ON-OFF) frames are given with grey-scale, with superposed contours of the OFF frames. 
J2000 celestial coordinates are given.}\label{results}
\end{figure}
\setcounter{figure}{13}
\begin{figure}
\caption{Continued.}
\end{figure}
\setcounter{figure}{13}
\begin{figure}
\caption{Continued.}
\end{figure}

\end{document}